\documentclass{article}
\usepackage{makeidx}
\usepackage{amsmath}
\usepackage{amsfonts}
\usepackage{amssymb}

\begin{document}

\title{Entanglement-breaking channels in infinite dimensions}
\author{A. S. Holevo \\
Steklov Mathematical Institute, Moscow}
\date{}
\maketitle

\begin{abstract} We give a representation for entanglement-breaking channels in
separable Hilbert space that generalizes the \textquotedblleft Kraus
decomposition with rank one operators\textquotedblright and use it
to describe the complementary channels. We also give necessary and
sufficient condition of entanglement-breaking for a general quantum
Gaussian channel. Application of this condition to one-mode channels
provides several new cases where the additivity conjecture holds in
full generality.
\end{abstract}

\section{Introduction}\label{0}

In the paper \cite{hsw} we gave a general integral representation
for separable states in the tensor product of infinite dimensional
Hilbert spaces and proved the structure theorem for the quantum
communication channels that are entanglement-breaking, which
generalizes the finite-dimensional result of Horodecki, Shor and
Ruskai \cite{R}.

In what follows $\mathcal{H}$ denotes separable Hilbert space;
$\mathfrak{T} ( \mathcal{H})$ -- the Banach space of trace-class
operators in $\mathcal{H}$, and $\mathfrak{S}(\mathcal{H})$ -- the
convex subset of all density operators $\rho $ in $\mathcal{H}$. We
shall also call them \textit{states } for brevity, having in mind
that a density operator $\rho $ uniquely determines a normal state
on the algebra $\mathfrak{B}(\mathcal{H})$ of all bounded operators
in $\mathcal{H}$ (see e. g. \cite{h0}). Equipped with the trace-norm
distance, $ \mathfrak{S}(\mathcal{H})$ is a complete separable
metric space.

A state $\rho \in \mathfrak{S}(\mathcal{H}_{1}\otimes \mathcal{H}_{2})$
 is called\textit{\ separable }if it belongs to the convex
closure of the set of all product states in $\mathfrak{S}(\mathcal{H}
_{1}\otimes \mathcal{H}_{2})$. It is shown in \cite{hsw} that separable
states are precisely those which admit a representation
\begin{equation}
\rho =\int_{\mathcal{X}}\rho _{1}(x)\otimes \rho _{2}(x)\pi (dx),
\end{equation}
where $\pi (dx)$ be a Borel probability measure and $\rho
_{j}(x),j=1,2,$ are Borel
$\mathit{\mathfrak{S}(\mathcal{H}_{j})}$-valued functions on some
complete separable metric space $\mathcal{X}.$

A \textit{channel} with input space $\mathcal{H}_{A}$ and output
space $ \mathcal{H}_{B}$ is bounded linear  completely positive
trace-preserving map $\Phi :$ $\mathfrak{ T
}(\mathcal{H}_{A})\rightarrow \mathfrak{T}(\mathcal{H}_{B})$.
Channel\textit{\ $ \Phi $ }is called\textit{\ entanglement-breaking
}if for arbitrary Hilbert space $\mathcal{H}_{R}$ and arbitrary
state $\rho \in \mathfrak{S}(\mathcal{ H }_{A}\otimes
\mathcal{H}_{R})$ the state $(\Phi \otimes \mathrm{Id} _{R})(\rho
)\in \mathfrak{S}(\mathcal{H}_{B}\otimes \mathcal{H}_{R})$, where
$\mathrm{Id}_{R}$ is the identity map in
$\mathfrak{T}(\mathcal{H}_{R})$, is separable.

It is shown in \cite{hsw} that the channel $\Phi $ is
entanglement-breaking if and only if there is a complete separable
metric space $\mathcal{X}$, a Borel
$\mathfrak{S}(\mathcal{H}_{B})$-valued function $\rho _{B}(x)$ and
an observable $M$ in $\mathcal{H}_{A}$ with the outcome set
$\mathcal{X}$ given by probability operator-valued measure (POVM)
$M(dx)$ (which is a measure on $ \mathcal{X}$ taking values in the
positive cone of $\mathfrak{B}(\mathcal{H} _{A})$, with
$M(\mathit{\mathcal{X}})=I$) such that
\begin{equation}
\Phi (\rho )=\int\limits_{\mathcal{X}}\rho _{B}(x)m_{\rho }(dx),  \label{thm}
\end{equation}
where $m_{\rho }(S)=\mathrm{Tr}\rho M(S)$ for all Borel subsets
$S\in\mathcal{B}(\mathcal{X})$. This gives a continual version of
the class of channels introduced in \cite {obzor} and can be
regarded as a generalization of a result in \cite{R} to infinite
dimensions.

In finite dimensions entanglement-breaking channels form a large
class in which the famous additivity conjecture for the classical
capacity holds as shown by Shor \cite {shor}. Generalization of this
property to infinite dimensions is by no means straightforward.
First, we define \textit{generalized ensemble} as arbitrary
probability distribution on the state space $\mathfrak{S}(\mathcal{
H}_{A})$ \cite{H-c-w-c}, \cite{H-Sh-2}. The average state of the
ensemble is given by the barycenter $\bar{\rho}_{\pi }=\int \rho \pi
(d\rho ).$ Let $ \mathcal{A}$ be an arbitrary subset of
$\mathfrak{S}(\mathcal{H}_{A})$, then the
$\mathcal{A}$-\textit{constrained }$\chi -$\textit{capacity} of the
channel $\Phi $ is defined as
\begin{equation*}
C_{\chi }(\Phi ,\mathcal{A})=\sup_{\pi :\bar{\rho}_{\pi }\in
\mathcal{A} }\int H\left( \Phi \lbrack \rho ];\Phi \left[
\bar{\rho}_{\pi }\right] \right) \pi (d\rho ),
\end{equation*}
where $H(\rho ;\sigma )$ is the quantum relative entropy. In case
the output entropy is finite on $\mathcal{A}$ this amounts to
\begin{equation}
C_{\chi }(\Phi ,\mathcal{A})=\sup_{\sigma \in \mathcal{A}}\left[
H\left( \Phi \left[ \sigma \right] \right) -\hat{H}_{\Phi }\left(
\sigma \right) \right] ,  \label{chic}
\end{equation}
where
\begin{equation}
\hat{H}_{\Phi }\left( \sigma \right) =\inf_{\pi :\bar{\rho}_{\pi }=\sigma
}\int H\left( \Phi \lbrack \rho ]\right) \pi (d\rho )  \label{convcl}
\end{equation}
is the convex closure of the output entropy $H\left( \Phi \lbrack
\rho ]\right).$ When $\mathcal{A=}\mathfrak{S}(\mathcal{H}_{A})$ and
$\pi $ runs through ordinary ensembles given by probability
distributions with finite supports, this reduces to the familiar
definition of (unconstrained) $\chi -$ capacity $C_{\chi }(\Phi )$.

Then, as shown in \cite{Sh-2}, for an entanglement-breaking channel $\Phi
_{1}$ and arbitrary channel $\Phi _{2}$ the additivity conjecture holds in
its strongest form
\begin{equation}
C_{\chi }(\Phi _{1}\otimes \Phi _{2},\mathcal{A}_{1}\otimes
\mathcal{A} _{2})=C_{\chi }(\Phi _{1},\mathcal{A}_{1})+C_{\chi
}(\Phi _{2},\mathcal{A} _{2}),  \label{hiad}
\end{equation}
where
\begin{equation*}
\mathcal{A}_{1}\otimes \mathcal{A}_{2}=\left\{ \rho \in
\mathfrak{S}( \mathcal{H}_{A_{1}}\otimes
\mathcal{H}_{A_{2}}):\mathrm{Tr}_{2}\rho \in
\mathcal{A}_{1},\mathrm{Tr}_{1}\rho \in \mathcal{A}_{2}\right\} .
\end{equation*}
Moreover, the convex closure is superadditive
\begin{equation}
\hat{H}_{\Phi _{1}\otimes \Phi _{2}}\left( \sigma _{12}\right) \geq
\hat{H} _{\Phi _{1}}\left( \sigma _{1}\right) +\hat{H}_{\Phi
_{2}}\left( \sigma _{2}\right)  \label{hiad1}
\end{equation}
for any state $\sigma _{12}$.

In applications the constraints of the form
\begin{equation}
\mathcal{A}(H,E)=\left\{ \rho \in
\mathfrak{S}(\mathcal{H}_{A}):\mathrm{Tr} \rho H\leq E\right\} ,
\label{2constraint}
\end{equation}
where $H$ is positive selfadjoint operator (typically the energy
operator, see \cite{H-c-w-c}) are of the major interest. In this
case one shows as in \cite {hs}, that (\ref{hiad}) implies
\begin{equation*}
C_{\chi }(\Phi ^{\otimes n},\mathcal{A}(H\otimes \dots \otimes
I+\dots I\otimes \dots \otimes H,nE))=nC_{\chi }(\Phi
,\mathcal{A}(H,E))
\end{equation*}
and hence
\begin{equation}
C(\Phi ,H,E)=C_{\chi }(\Phi ,\mathcal{A}(H,E))=\sup_{\pi
:\mathrm{Tr}\bar{ \rho}_{\pi }H\leq E}\int H\left( \Phi \lbrack \rho
];\Phi \left[ \bar{\rho} _{\pi }\right] \right) \pi (d\rho )
\label{clca}
\end{equation}
for any entanglement-breaking channel $\Phi ,$ where $C(\Phi ,H,E)$
is the classical capacity of the channel $\Phi $ with the input
energy constraint as defined in \cite{H-c-w-c}.

The paper has two self-consistent parts. In part I (Sec. \ref{1},
\ref{2}) we give another representation for entanglement-breaking
channels in separable Hilbert space, that generalizes the
\textquotedblleft Kraus decomposition with rank one
operators\textquotedblright. We also find complementary channels and
remark that coherent information for anti-degradable channel is
always non-positive.

Part II  (Sec. \ref{3}-\ref{6}) is devoted to Gaussian
entanglement-breaking channels. We give necessary and sufficient
condition of entanglement-breaking for a general quantum Gaussian
channel. Application of this condition to one-mode channels provides
several new cases where the additivity conjecture holds in the full
generality.

\section{ A representation for entanglement-breaking
channels}\label{1}

 Here we further specify the formula (\ref{thm}), by employing the representation for POVM from
\cite{3inst}. Basing on this specification, we give explicit
description of the Stinespring isometry for the channel $\Phi $ and
of the complementary channel.

\textbf{Lemma 1.} (Radon-Nikodym theorem for POVM) \textit{For every
POVM $M$ on $\mathcal{X}$, there exist a positive $\sigma $-finite
measure $\mu $ on $\mathcal{X}$, a dense domain $\mathcal{D\subset
H}$, and a countable family of functions $ x\rightarrow a_{k}(x)$
such that for almost all $x$, $a_{k}(x)$ are linear functionals on
$\mathcal{D}$, satisfying}
\begin{equation}
\int_{\mathcal{X}}\sum_{k}\left\vert \left\langle a_{k}(x)|\psi
\right\rangle \right\vert ^{2}\mu (dx)=\Vert \psi \Vert ^{2};~~~\psi
\in \mathcal{D},  \label{2.4}
\end{equation}
\textit{and}
\begin{equation}
\left\langle \psi |M(S)\psi \right\rangle
=\int_{S}\sum_{k}\left\vert \left\langle a_{k}(x)|\psi \right\rangle
\right\vert ^{2}\mu (dx);~~~\psi \in \mathcal{D}.  \label{2.5}
\end{equation}

Note that $a_{k}(x)$ are in general unbounded functionals defined
only on $\mathcal{D}$, nevertheless we find it convenient to
continue use of the ``bra-ket'' notation for such functionals.

\textit{Proof:} We follow the proof of Radon-Nikodym theorem for
instruments from \cite{3inst}. By Naimark's theorem \cite{h0}, there
exist a Hilbert space $\mathcal{K}$, and sharp observable $E$ given
by spectral measure $
\left\{E(S);S\in\mathcal{B}(\mathcal{X})\right\} $ in $\mathcal{K}$,
and an isometry $W$ from $\mathcal{H}$ to $\mathcal{K}$ such that
\begin{equation}
M(S)=W^{\ast }E(S)W,\quad S\in \mathcal{B}(\mathcal{X}).
\label{2.1}
\end{equation}
According to von Neumann's spectral theorem, $\mathcal{K}$ can be
decomposed into the direct integral of Hilbert spaces
\begin{equation}
\mathcal{K}=\int_{\mathcal{X}}\oplus \mathcal{H}(x)\mu (dx)
\label{2.6}
\end{equation}
with respect to some positive $\sigma $-finite measure $\mu $,
diagonalizing the spectral measure $E$:
\begin{equation}
E(S)\phi =\int_{S}\oplus \phi (x)\mu (dx),  \label{2.7}
\end{equation}
where $\phi (x)\in \mathcal{H}(x)$ are the components of the vector
$\phi \in \mathcal{K}$ in the decomposition (\ref{2.6}). Let us fix
a measurable field of orthonormal bases $\{e_{k,x}\}$ in the direct
integral  (\ref{2.6}) and denote $\phi _{k}(x)=\left\langle
e_{k,x}|\phi (x)\right\rangle $, where the inner product is in
$\mathcal{H}(x)$ (note that $\phi _{k}(x)$ are defined $ \mu
$-almost everywhere).

Let $\psi \in \mathcal{H}$, then the decomposition of the vector
$W\psi \in \mathcal{K}$ reads
\begin{equation*}
W\psi =\int_{\mathcal{X}}\oplus \sum_{k}(W\psi )_{k}e_{k,x}\mu (dx),
\end{equation*}
where
\begin{equation}
\int_{\mathcal{X}}\sum_{k}\left\vert (W\psi )_{k}(x)\right\vert
^{2}\mu (dx)=\Vert \psi \Vert ^{2};~~\psi \in \mathcal{H},
\label{2.8}
\end{equation}
since $W$ is isometric. Since $W$ is linear, we have for $\mu
$-almost all $ x $
\begin{equation}
(W(\sum_{j}\lambda _{j}\psi _{j}))_{k}(x)=\sum_{j}\lambda _{j}(W\psi
_{j})_{k}(x),  \label{2.9}
\end{equation}
where $\{\psi _{j}\}\subset \mathcal{H}$ is a fixed system of
vectors, and $ \lambda _{j}$ are complex numbers, only finite number
of which are non-zero.

Let us now fix an orthonormal basis $\{\psi _{j}\}\subset
\mathcal{H}$ and let $\mathcal{D}=\mbox{lin}\{\psi _{j}\}$ be its
linear span. We define linear functionals $a_{k}(x)$ on
$\mathcal{D}$ by the relation
\begin{equation*}
\left\langle a_{k}(x)|\sum_{j}\lambda _{j}\psi _{j}\right\rangle
=\sum_{j}\lambda _{j}(W\psi _{j})_{k}^{0}(x),
\end{equation*}
where $(W\psi _{j})_{k}^{0}$ is a fixed representative of the
equivalence class $(W\psi _{j})_{k}$. Then by (\ref{2.9}), for any
fixed $\psi \in \mathcal{D}$ there is a subset $\mathcal{X_{\psi
}\subset X}$, such that $ \mu (\mathcal{X\setminus X}_{\psi })=0$
and
\begin{equation}
\left\langle a_{k}(x)|\psi \right\rangle =(W\psi )_{k}(x),\mbox{ for
all}\,k, \mbox{ all }x\in \mathcal{X_{\psi }}.  \label{2.10}
\end{equation}
Combining (\ref{2.7}) and (\ref{2.10}), we obtain (\ref{2.5}). The
normalization condition (\ref{2.4}) follows from (\ref{2.8}) and
(\ref{2.10} ).$\square $

\bigskip \textbf{Theorem 1.} \textit{For any entanglement-breaking channel $\Phi $
there exist a complete separable metric space $\mathcal{Y}$, a
positive $ \sigma $-finite measure $\nu $ on $\mathcal{Y}$, a dense
domain $\mathcal{ D\subset H}_{A}$, a measurable function
$y\rightarrow a(y)$, defined for almost all $y$, such that $a(y)$
are linear functionals on $\mathcal{D}$, satisfying}
\begin{equation}
\int_{\mathcal{Y}}\left\vert \left\langle a(y)|\psi \right\rangle
\right\vert ^{2}\nu (dy)=\Vert \psi \Vert ^{2};~~~\psi \in
\mathcal{D}, \label{nar}
\end{equation}
\textit{and a measurable family of unit vectors $y\rightarrow b(y)$
in $\mathcal{H} _{B}$ such that}
\begin{equation}
\Phi \lbrack \rho ]=\int\limits_{\mathcal{Y}}\left\vert b(y)\rangle
\langle b(y)\right\vert \left\vert \left\langle a(y)|\psi
\right\rangle \right\vert ^{2}\nu (dy)~~\mathit{for} \rho
=\left\vert \psi \rangle \langle \psi \right\vert \mathit{ with
}\,\psi \in \mathcal{D}.  \label{repa}
\end{equation}

 This is infinite-dimensional analog of the \textquotedblleft Kraus
decomposition with rank one operators\textquotedblright\ from
\cite{R} with the difference that it is in general continual and the
Kraus operators $ V(y)=\left\vert b(y)\rangle \langle
a(y)\right\vert $ are unbounded and only densely defined. As shown
in \cite{hsw}, there are entanglement-breaking channels for which
the usual Kraus decomposition with bounded rank one operators does
not exist.

\textit{Proof.} By making the spectral decomposition of the density
operators $\rho _{B}(x)=\sum_{l}\lambda _{l}(x)\left\vert
b_{l}(x)\rangle \langle b_{l}(x)\right\vert $ in (\ref{thm}) and
using lemma 1, we find
\begin{equation}
\Phi \lbrack \rho
]=\int\limits_{\mathcal{X}}\sum\limits_{k,l}\lambda
_{l}(x)\left\vert b_{l}(x)\rangle \langle b_{l}(x)\right\vert
\left\vert \left\langle a_{k}(x)|\psi \right\rangle \right\vert
^{2}\mu (dx);~~~\psi \in \mathcal{D}.  \label{repb}
\end{equation}
Let $\mathcal{Y}$ be the space of triples $y=(x,k,l),$ with
naturally defined metric and the countably finite measure defined by
\begin{equation*}
\nu (S\times k\times l)=\int\limits_{S}\lambda _{l}(x)\mu (dx).
\end{equation*}
Define $a(y)=a_{k}(x)$ and $b(y)=b_{l}(x),$ then (\ref{nar}) follows
from (\ref{2.4}) and (\ref{repa}) -- from (\ref{repb}).$\square $

\section{Remarks on complementary channels}\label{2}

In general, the Stinespring representation
\begin{equation*}
\Phi \lbrack \rho ]=\text{Tr}_{E}V\rho V^{\ast }
\end{equation*}
holds in the infinite dimensional case for arbitrary channel, where
$\mathcal{H}_{E}$ is environment space and
$V:\mathcal{H}_{A}\rightarrow \mathcal{H}_{B}\otimes
\mathcal{H}_{E}$ is an isometry. The complementary channel is
defined as
\begin{equation*}
\tilde{\Phi}[\rho ]=\mathrm{Tr}_{B}V\rho V^{\ast }.
\end{equation*}
If there exists a channel $T:$
$\mathfrak{T}(\mathcal{H}_{B})\rightarrow
\mathfrak{T}(\mathcal{H}_{E})$ such that $\tilde{\Phi}=T\circ \Phi
,$ then $ \Phi $ is degradable \cite{dev}, and if $\Phi =T^{\prime
}\circ \tilde{\Phi}$ for some channel $T^{\prime }:$
$\mathfrak{T}(\mathcal{H}_{E})\rightarrow
\mathfrak{T}(\mathcal{H}_{B}),$ then $\Phi $ is anti-degradable
\cite{car}.

\textbf{Proposition 1.} \textit{If }$\Phi $\textit{\ is degradable
(anti-degradable) channel then }$I_{c}(\rho ,\Phi )\geq 0$\textit{\
(resp. }$ I_{c}(\rho ,\Phi )\leq 0)$\textit{\ for arbitrary density
operator }$\rho $ \textit{such that} $H(\Phi \lbrack \rho ])<\infty
,H(\tilde{\Phi}[\rho ])<\infty .$

\textit{Proof.} As noticed in  \cite{sch}, there is a formula which
relates the coherent information and the $\chi -$ quantity giving
the upper bound for the classical information. Namely, for arbitrary
pure-state decomposition $\rho =\sum_{j}\pi _{j}\rho _{j}$
\begin{equation*}
I_{c}(\rho ,\Phi )=\chi_B-\chi_E,
\end{equation*}
where
\begin{equation*}
\chi_B=H(\Phi \lbrack \rho ])-\sum_{j}\pi _{j}H(\Phi [\rho _{j}])
\end{equation*}
and similarly for $\chi_E$. But $\chi_B=\sum_{j}\pi _{j} H(\Phi
\lbrack \rho _{j}];\Phi \lbrack \rho ])$, whence
\begin{equation*} I_{c}(\rho ,\Phi
)=\sum_{j}\pi _{j}\left[ H(\Phi \lbrack \rho _{j}];\Phi \lbrack \rho
])-H( \tilde{\Phi}[\rho _{j}];\tilde{\Phi}[\rho ])\right].
\end{equation*}
The assertion then follows from the monotonicity of the relative
entropy and the definition of (anti-)degradable channel. $\square $

As observed in \cite{car}, anti-degradable channels have zero
quantum capacity $Q(\Phi)$. Proposition 1 provides a short proof of
this statement. Let $\Phi $ be anti-degradable, then such is $\Phi
^{\otimes n},$ hence $I_{c}(\rho ,\Phi^{\otimes n} )\le 0.$ Then the
coding theorem for the quantum capacity implies
\begin{equation*}
Q(\Phi)=\lim_{n\to\infty}n^{-1}\sup_{\rho}I_{c}(\rho ,\Phi^{\otimes
n} )=0.
\end{equation*}

For the entanglement-breaking channel (\ref{repa}) we introduce the
environment space $\mathcal{H}_{E}=L^{2}(\nu ),$ and define the
operator $V:\mathcal{D\rightarrow H}_{B}\otimes \mathcal{H}_{E}$ by
the formula
\begin{equation*}
\left( V\psi \right) (y)=|b(y)\rangle \left\langle a(y)|\psi
\right\rangle .
\end{equation*}
Then $V$ is isometric by (\ref{nar}) and hence uniquely extends to
the isometry $\mathcal{H}_{A}\mathcal{\rightarrow H}_{B}\otimes
\mathcal{H}_{E}.$ It is the Sinespring isometry for channel $\Phi $
namely
\begin{equation*}
\mathrm{Tr}_{E}V\rho V^{\ast }=\Phi \lbrack \rho ],\quad
\text{~\textrm{for} } \rho =\left\vert \psi \rangle \langle \psi
\right\vert \text{ with }\psi \in \mathcal{D}
\end{equation*}
by (\ref{repa}). On the other hand, $\mathrm{Tr}_{B}V\rho V^{\ast }$
is the integral operator in $\mathcal{H}_{E}=L^{2}(\nu )$ defined by
the kernel
\begin{equation*}
\sigma _{\rho }(y_{2},y_{1})=\left\langle
b(y_{2})|b(y_{1})\right\rangle \left\langle a(y_{1})|\psi
\right\rangle \overline{\left\langle a(y_{2})|\psi \right\rangle },
\end{equation*}
which completely describes the complementary channel $\tilde{\Phi}$.

\textbf{Example.} Let $\mathcal{H}_{A}=L^{2}(\nu ),$
$\mathcal{D=}C(\mathcal{ Y})\cap L^{2}(\nu ),$ then the relation
\begin{equation*}
\Phi \lbrack \rho ]=\int\limits_{\mathcal{Y}}\left\vert b(y)\rangle
\langle b(y)\right\vert \left\vert \psi (y)\right\vert ^{2}\nu
(dy);~~\text{~\textrm{ \ for} }\rho =\left\vert \psi \rangle \langle
\psi \right\vert \text{ with } \psi \in \mathcal{D}
\end{equation*}
defines entanglement-breaking channel. This extends to
\begin{equation*}
\Phi \lbrack \rho ]=\int\limits_{\mathcal{Y}}\left\vert b(y)\rangle
\langle b(y)\right\vert \rho (y,y)dy,
\end{equation*}
where $\rho (y_{2},y_{1})$ is the kernel of the integral operator
$\rho $ in $L^{2}(\nu)$ (for which the diagonal value $\rho (y,y)$
is unambiguously defined \cite{vuz}). The output of the
complementary channel is the integral operator in
$\mathcal{H}_{E}=L^{2}(\nu)$ with the kernel
\begin{equation*}
\sigma _{\rho }(y_{2},y_{1})=\left\langle
b(y_{2})|b(y_{1})\right\rangle \rho (y_{2},y_{1}).
\end{equation*}

In finite dimensions \textit{every entanglement-breaking channel is
anti-degradable}\footnote{We are indebted to M.-B.Ruskai for this
observation \cite{crs}.}. In infinite dimensions, use representation
(\ref{repa}) and define the entanglement-breaking channel $T^{\prime
}:$ $ \mathfrak{T}(\mathcal{H}_{E})\rightarrow
\mathfrak{T}(\mathcal{H}_{B})$ by the formula
\begin{equation*}
T^{\prime }[\sigma ]=\int\limits_{\mathcal{Y}}\left\vert b(y)\rangle
\langle b(y)\right\vert \sigma (y,y)\nu (dy),
\end{equation*}
then ~$\Phi \lbrack \rho ]=T^{\prime }[\tilde{\Phi}[\rho ]]$ for
$\rho =\left\vert \psi \rangle \langle \psi \right\vert $ with $\psi
\in \mathcal{D }$, hence the assertion follows.

\section{Gaussian observables}\label{3}

In what follows we shall consider real vector space $Z$ equipped
with different bilinear forms $\alpha ,\Delta ,\dots .$ For
concreteness and convenience of notation we shall consider $Z$ as
the space of column vectors with components in $\mathbb{R}.$ Then
the forms are given by matrices which we denote by the same letter,
e. g. $\alpha (w,z)=w^{T}\alpha z,$ where $^{T} $ denotes
transposition. If $\alpha $ is an inner product on $Z,$ then $
(Z,\alpha )$ is an Euclidean space, while if $\Delta $ is a
nondegenerated skew-symmetric form then $(Z,\Delta )$ is a
symplectic space. We call symplectic space $(Z,\Delta )$ standard if
the commutation matrix, corresponding to the symplectic form $\Delta
(z,z^{\prime }),$ is $\Delta = \mathrm{diag}\left[
\begin{array}{cc}
0 & -1 \\
1 & 0
\end{array}
\right] .$ Let $2n=\dim Z,$ and let $d^{2n}z$ denote element of symplectic
volume in $(Z,\Delta ).$ Symplectic Fourier transform and its converse are
given by
\begin{equation*}
\hat{f}(w)=\int \mathrm{e}^{i\Delta (w,z)}f(z)d^{2n}z;\quad
f(z)=\left( 2\pi \right) ^{-2n}\int \mathrm{e}^{-i\Delta
(w,z)}\hat{f}(w)d^{2n}w.
\end{equation*}

Quantization on a symplectic space $(Z,\Delta )$ is given by (irreducible)
Weyl system in a Hilbert space $\mathcal{H}$, which is a family of unitary
operators $\left\{ W(z),z\in Z\right\} $ satisfying the canonical
commutation relations, one of the equivalent forms of which is
\begin{equation}
W(z)^{\ast }W(w)W(z)=\exp \left( i\Delta (w,z)\right) W(w).  \label{www}
\end{equation}
By Stone's theorem, $W(z)=\exp \left( iRz\right) $, where $R$ is the row
vector of selfadjoint operators in $\mathcal{H}$ satisfying the Heisenberg
commutation relations, see \cite{h0}, Ch. V, for detail.

Assume we have two symplectic spaces $Z_{A},Z_{B}$ with the corresponding
Weyl systems in Hilbert spaces $\mathcal{H}_{A},\mathcal{H}_{B}.$ Let $M$ an
observable in $\mathcal{H}_{A}$ with the outcome set $Z_{B},$ given by
positive operator-valued measure $M(d^{2n}z)$. In what follows we skip the
index $_{B}$ so that $Z=Z_{B}$ etc. It is completely determined by the
\textit{operator characteristic function }\cite{h2}
\begin{equation*}
\phi _{M}(w)=\int \mathrm{e}^{i\Delta (w,z)}M(d^{2n}z).
\end{equation*}
A function $\phi (w)$ is characteristic function of an observable if and
only if it satisfies the following conditions:

\begin{enumerate}
\item $\phi (0)=I;$

\item $\phi (w)$ is weakly continuous at $w=0;$

\item for any choice of a finite subset $\left\{ w_{j}\right\} \subset Z_{B}$
the block matrix with operator entries $\phi (w_{j}-w_{k})$ is nonnegative
definite.
\end{enumerate}

In the case $\left\Vert \phi _{M}(w)\right\Vert $ is integrable on $Z_{B},$
measure $M(d^{2n}z)$ has density $p_{M}(z),$ $M(d^{2n}z)=p_{M}(z)d^{2n}z,$
which is a.e. defined function on $Z_{B},$ taking values in the cone of
bounded positive operators in $\mathcal{H}_{A}$, such that $\int
p_{M}(z)d^{2n}z=I.$ Moreover,
\begin{equation}
p_{M}(z)=\left( 2\pi \right) ^{-2n}\int \mathrm{e}^{-i\Delta (w,z)}\phi
_{M}(w)d^{2n}w.  \label{invert}
\end{equation}

Observable $M$ will be called \textit{Gaussian }(\textit{canonical
}in \cite {h2}, \cite{h0}) if its operator characteristic function
has the form
\begin{equation}
\phi _{M}(w)=W_{A}(Kw)\exp \left( -\frac{1}{2}\mu (w,w)\right) =\exp \left(
iR_{A}Kw-\frac{1}{2}\mu (w,w)\right) ,  \label{ocf}
\end{equation}
where $K:Z_{B}\rightarrow Z_{A}$ is a linear operator and $\mu $ is
a bilinear form on $Z_{B}.$ A necessary and sufficient condition for
(\ref{ocf} ) to define an observable is the matrix inequality
\begin{equation}
\mu \geq \frac{i}{2}K^{T}\Delta _{A}K.  \label{mudelta}
\end{equation}
Indeed, (\ref{ocf}) apparently satisfies conditions 1,2 and
(\ref{mudelta}) is equivalent to the condition 3 since for an
operator function given by ( \ref{ocf})
\begin{eqnarray*}
\phi (w_{j}-w_{k}) &=&W(Kw_{k})^{\ast }W(Kw_{j})\exp \left[
-\frac{i}{2}
\Delta (Kw_{j},Kw_{k})-\frac{1}{2}\mu (w_{j}-w_{k},w_{j}-w_{k})\right]  \\
&=&C_{k}^{\ast }C_{j}\exp \left[ \mu (w_{j},w_{k})-\frac{i}{2}\Delta
(Kw_{j},Kw_{k})\right] ,
\end{eqnarray*}
where $C_{j}=W(Kw_{j})\exp \left[ -\frac{1}{2}\mu
(w_{j},w_{j})\right] ,$ and nonnegative definiteness of matrices
with scalar entries $\mu (w_{j},w_{k})- \frac{i}{2}\Delta
(Kw_{j},Kw_{k}),$ with arbitrary choice of finite subset $ \left\{
w_{j}\right\} ,$ is equivalent to that for $\exp \left[ \mu
(w_{j},w_{k})-\frac{i}{2}\Delta (Kw_{j},Kw_{k})\right] $ (see
\cite{h0}, proof of Theorem 5.1, Ch. V).

Observable $M$ is sharp if and only if $\mu =0,$ in which case it is the
spectral measure of commuting selfadjoint operators $R_{A}K.$ In a sense
opposite is the following case:

\textbf{Proposition 2.} \textit{Let $\mu $ be nondegenerate and $K$
invertible, then observable $M$ has operator density given by}
\begin{equation*}
p_{M}(z)=W_{A}(K^{\prime -1}z)\sigma _{A}W_{A}(K^{\prime -1}z)^{\ast
}\frac{1}{\left( 2\pi \right) ^{n}\left\vert \det K\right\vert },
\end{equation*}
\textit{where $\sigma _{A}$ is Gaussian state with zero mean and
correlation function $\mu (K^{-1}z,K^{-1}z^{\prime })$ and
$K^{\prime }=\Delta ^{-1}K^{T}\Delta $ is symplectic transpose.}

\textit{Proof.} Since $\mu $ is nondegenerate, $\left\Vert \phi
_{M}(w)\right\Vert =\exp \left( -\frac{1}{2}\mu (w,w)\right) $ is
integrable, and applying (\ref{invert}) we get
\begin{eqnarray*}
p_{M}(z) &=&\left( 2\pi \right) ^{-2n}\int \mathrm{e}^{-i\Delta
(w,z)}W_{A}(Kw)\exp \left( -\frac{1}{2}\mu (w,w)\right) d^{2n}w \\
&=&\left( 2\pi \right) ^{-2n}\int \mathrm{e}^{i\Delta (u,K^{\prime
-1}z)}W_{A}(-u)\exp \left( -\frac{1}{2}\mu (K^{-1}u,K^{-1}u)\right)
\frac{
d^{2n}u}{\left\vert \det K\right\vert } \\
&=&\left( 2\pi \right) ^{-2n}\int W_{A}(K^{\prime
-1}z)W_{A}(-u)W_{A}(K^{\prime -1}z)^{\ast }\exp \left( -\frac{1}{2}\mu
(K^{-1}u,K^{-1}u)\right) \frac{d^{2n}u}{\left\vert \det K\right\vert } \\
&=&\frac{1}{\left( 2\pi \right) ^{n}\left\vert \det K\right\vert }
W_{A}(K^{\prime -1}z)\left[ \int W_{A}(-u)\exp \left(
-\frac{1}{2}\mu (K^{-1}u,K^{-1}u)\right) \frac{d^{2n}u}{\left( 2\pi
\right) ^{n}}\right] W_{A}(K^{\prime -1}z)^{\ast }.
\end{eqnarray*}
Here we used change of variable $u=-Kw$ in the second line, the
relation ( \ref{www}) in the third line and in the fourth line the
term in squared brackets is the Weyl transform of the characteristic
function of the state $ \sigma _{A}.$ $\square $

Let us describe an explicit construction of Naimark's dilation of observable
$M$ in the spirit of \cite{h0}, Prop. 5.1, Ch. II.

\textbf{Proposition 3.} \textit{Assume the condition (\ref{mudelta})
holds, then there exist Bosonic system $\mathcal{H}_{C}$ with
canonical observables $ R_{C}$ such that $\mathcal{H}_{B}\subseteq
\mathcal{H}_{A}\otimes \mathcal{H} _{C}$ and Gaussian state $\rho
_{C}\in \mathfrak{S}(\mathcal{H}_{C})$ for which}
\begin{equation}
M(S)=\mathrm{Tr}_{C}\left( I_{A}\otimes \rho _{C}\right)
E_{AC}(S),\quad S\subseteq Z_{B},  \label{nai}
\end{equation}
\textit{where $E_{AC}$ is a sharp observable in
$\mathcal{H}_{A}\otimes \mathcal{H} _{C}$ given by the joint
spectral measure of commuting selfadjoint operators}
\begin{equation}
X_{B}=\Delta _{B}^{-1}\left( R_{A}K+R_{C}K_{C}\right) ^{T},  \label{selfa}
\end{equation}
\textit{where $K_{C}:Z_{B}\rightarrow Z_{C}$ is operator such that}
\begin{equation}
K_{C}^{T}\Delta _{C}K_{C}=-K^{T}\Delta _{A}K.  \label{comuta}
\end{equation}

\textit{Proof.} The condition (\ref{comuta}) means that
$K_{C}^{T}\Delta _{C}K_{C}+K^{T}\Delta _{A}K=0,$ that is
commutativity of operators (\ref {selfa}). By adapting the proof of
Prop. 8.1 from Ch. VI of \cite{h0}, we obtain a symplectic space
$(Z_{C},\Delta _{C}),$ operator $ K_{C}:Z_{B}\rightarrow Z_{C}$ and
an inner product in $Z_{C},$ given by symmetric matrix $\alpha
_{C}\geq \frac{i}{2}\Delta _{C}$ such that (\ref {comuta}) holds
along with
\begin{equation*}
\quad K_{C}^{T}\alpha _{C}K_{C}=\mu .
\end{equation*}
Then the characteristic function of $E_{AC}$ is
\begin{eqnarray*}
\phi _{E_{AC}}(w_{B}) &=&\int \exp [i\Delta
(w_{B},z_{B})]E_{AC}\left(
d^{2n}z_{B}\right)  \\
&=&\exp iw_{B}^{T}\Delta _{B}X_{B}=\exp i\left( R_{A}K+R_{C}K_{C}\right)
w_{B} \\
&=&W_{A}(Kw_{B})W_{C}(K_{C}w_{B}),
\end{eqnarray*}
whence
\begin{eqnarray*}
\mathrm{Tr}_{C}\left( I_{A}\otimes \rho _{C}\right) \phi
_{E_{AC}}(w_{B}) &=&W_{A}(Kw_{B})\exp \left( -\frac{1}{2}\alpha
_{C}(K_{C}w_{B},K_{C}w_{B})\right)  \\
&=&W_{A}(Kw_{B})\exp \left( -\frac{1}{2}\mu (w_{B},w_{B})\right) =\phi
_{M}(w_{B}),
\end{eqnarray*}
and (\ref{nai}) follows.$\square $

\section{Gaussian entanglement-breaking channels}\label{4}

Recall that characteristic function of a state $\rho $ is given by
\begin{equation*}
\varphi (z)=\mathrm{Tr}\rho W(z).
\end{equation*}
Channel $\Phi :\mathfrak{T}(\mathcal{H}_{A})\rightarrow \mathfrak{T}(
\mathcal{H}_{B})$ transforming states according to the rule
\begin{equation}
\varphi _{B}(z_{B})=\varphi _{A}(Kz_{B})f(z_{B}),  \label{linbos}
\end{equation}
where $K$ is a linear map between output and input symplectic spaces
$ (Z_{B},\Delta _{B})$, $(Z_{A},\Delta _{A})$ and $f$ is a complex
function, is called \textit{linear Bosonic}. Necessary and
sufficient condition on $f$ is nonnegative definiteness of matrices
with scalar entries
\begin{equation*}
f(w_{j}-w_{k})\exp \left[ -\frac{i}{2}\Delta (w_{j},w_{k})+\frac{i}{2}\Delta
(Kw_{j},Kw_{k})\right] ,
\end{equation*}
with arbitrary choice of finite subset $\left\{ w_{j}\right\} \subset Z_{B}$
\cite{demoen}, \cite{hw}.

If, additionally, $f$ is a Gaussian characteristic function, the channel is
\textit{Gaussian}. Thus for Gaussian channel, transformation of states is
described as
\begin{equation}
\varphi _{B}(z_{B})=\varphi _{A}(Kz_{B})\exp \left[ im(z_{B})-\frac{1}{2}
\alpha (z_{B},z_{B})\right] .  \label{gausch}
\end{equation}
The triple $(K,m,\alpha )$ is called parameters of the Gaussian channel.
Without loss of generality we assume $m\equiv 0.$ Necessary and sufficient
condition on the parameters of Gaussian channel is
\begin{equation}
\alpha \geq \frac{i}{2}\left[ \Delta _{B}-K^{T}\Delta _{A}K\right] .
\label{nis}
\end{equation}
This follows from the condition for the general linear Bosonic channel
applied to Gaussian $f$ similarly to the proof of (\ref{mudelta}).
Importance of this condition in the matrix form was emphasized in \cite{ew}.

\textbf{Theorem 2.} \textit{Let $\Phi $ be quantum Gaussian channel
with parameters $ (K,0,\alpha ).$ It is entanglement-breaking if and
only if $\alpha $ admits decomposition}
\begin{equation}
\alpha =\nu +\mu ,\quad \mathrm{where}\quad \nu \geq \frac{i}{2}\Delta
_{B},\quad \mu \geq \frac{i}{2}K^{T}\Delta _{A}K.  \label{nsc}
\end{equation}
\textit{In this case $\Phi $ has the representation}
\begin{equation}
\Phi \lbrack \rho ]=\int_{Z_{B}}W(z)\sigma _{B}W(z)^{\ast }m_{\rho
}(d^{2n}z),  \label{gebreak}
\end{equation}
\textit{where $\sigma _{B}$ is Gaussian state with parameters
$(0,\nu ),$ and $ m_{\rho }(S)=\mathrm{Tr}\rho M_{A}(S),S\subseteq $
$Z_{B},$ is the probability distribution of the Gaussian observable
$M_{A}$ with characteristic function (\ref{ocf}).}

\textit{Proof.} First, assume $\alpha $ admits the decomposition and
consider the channel defined by (\ref{gebreak}); we have to show that
\begin{equation}
\Phi ^{\ast }[W_{B}(w)]=W_{A}(Kw)\exp \left[ -\frac{1}{2}\alpha (w,w)\right]
.  \label{gebraek}
\end{equation}
Indeed, for arbitrary $\rho $
\begin{eqnarray}
\mathrm{Tr}\rho \Phi ^{\ast }[W_{B}(w)] &=&\mathrm{Tr}\Phi \lbrack \rho
]W_{B}(w)=\int_{Z_{B}}\mathrm{Tr}W_{B}(z)\sigma _{B}W_{B}(z)^{\ast
}W_{B}(w)m_{\rho }(d^{2n}z)  \notag \\
&=&\int_{Z_{B}}\mathrm{Tr}\sigma _{B}W_{B}(z)^{\ast }W_{B}(w)W_{B}(z)m_{\rho
}(d^{2n}z)  \notag \\
&=&\mathrm{Tr}\sigma _{B}W_{B}(w)\int_{Z_{B}}\exp \left[ i\Delta (w,z)\right]
m_{\rho }(d^{2n}z)  \notag \\
&=&\exp \left[ -\frac{1}{2}\nu (w,w)\right] \mathrm{Tr}\rho \phi _{M_{A}}(w)
\label{line4} \\
&=&\mathrm{Tr}\rho W_{A}(Kw)\exp \left[ -\frac{1}{2}\nu (w,w)-\frac{1}{2}\mu
(w,w)\right] ,  \notag
\end{eqnarray}
whence (\ref{gebraek}) follows.

Conversely, let $\Phi $ be Gaussian and entanglement-breaking. We
will use Gaussian version of the proof from \cite{hsw}, generalizing
the Choi-Jamiolkowski correspondence to infinite-dimensional
channels. Fix a Gaussian state $\rho _{A}$ in
$\mathfrak{S}(\mathcal{H}_{A})$ of full rank and let
$\{|e_{i}\rangle \}_{i=1}^{+\infty }$ be the basis of eigenvectors
of $\rho _{A}$ with the corresponding (positive) eigenvalues
$\{\lambda _{i}\}_{i=1}^{+\infty }$. Consider the unit vector
\begin{equation*}
|\Omega \rangle =\sum_{i=1}^{+\infty }\sqrt{\lambda
_{i}}|e_{i}\rangle \otimes |e_{i}\rangle
\end{equation*}
in the space $\mathcal{H}_{A}\otimes \mathcal{H}_{A}$, then $|\Omega \rangle
\langle \Omega |$ is Gaussian purification of $\rho _{A}.$ Since $\Phi $ is
entanglement-breaking, the Gaussian state
\begin{equation}
\rho _{AB}=(\mathrm{Id}_{A}\otimes \Phi )\left[ |\Omega \rangle \langle
\Omega |\right]
\end{equation}
in $\mathfrak{S}(\mathcal{H}_{A}\otimes \mathcal{H}_{B})$ is
separable. As follows from the proof of Proposition 1 in \cite{ww},
this implies representation
\begin{equation*}
\rho _{AB}=\int\limits_{Z_{A}}\int\limits_{Z_{B}}W_{A}(z_{A})\sigma
_{A}W_{A}(z_{A})^{\ast }\otimes W_{B}(z_{B})\sigma _{B}W_{B}(z_{B})^{\ast
}P(d^{2m}z_{A}d^{2n}z_{B}),
\end{equation*}
where $\sigma _{A},\sigma _{B}$ are Gaussian states and $P$ is a Gaussian
probability distribution. One then shows as in \cite{hsw} that the relation
\begin{eqnarray*}
&&M_{A}(S) \\
&=&\rho _{A}^{-1/2}\left[
\int\limits_{Z_{A}}\int\limits_{S}\overline{ W_{A}(z_{A})\sigma
_{A}W_{A}(z_{A})^{\ast }}\otimes W_{B}(z_{B})\sigma
_{B}W_{B}(z_{B})^{\ast }P(d^{2m}z_{A}d^{2n}z_{B})\,\right] \rho
_{A}^{-1/2},
\end{eqnarray*}
where bar means complex conjugation in the basis of eigenvectors of
$\rho _{A},$ defines an observable on Borel subsets $S\subseteq
Z_{B},$ and the representation (\ref{gebreak}) holds for the channel
$\Phi $ with these $ M_{A}$ and $\sigma _{B}.$ Let us denote $\nu $
the correlation function of the state $\sigma _{B};$ without loss of
generality we can assume its mean is zero. It remains to show that
$M_{A}$ is Gaussian observable with the characteristic function
(\ref{ocf}) where $\mu =\alpha -\nu .$ But from (\ref {line4})
\begin{equation*}
\Phi ^{\ast }[W_{B}(w)]=\exp \left[ -\frac{1}{2}\nu (w,w)\right] \phi
_{M_{A}}(w)
\end{equation*}
for any channel $\Phi $ with the representation (\ref{gebreak}), whence
taking into account (\ref{gebraek}), we indeed get (\ref{ocf}) with $\mu
=\alpha -\nu .\square $

A necessary condition for the decomposability (\ref{nsc}) and hence for the
channel to be entanglement breaking is
\begin{equation}
\alpha \geq \frac{i}{2}\left( \Delta _{B}\pm K^{T}\Delta _{A}K\right) .
\label{ppt}
\end{equation}
In general this condition implies that for any input Gaussian state of the
channel $\mathrm{Id}_{A}\otimes \Phi ,$ the output has positive partial
transpose. Indeed, this channel transforms the correlation matrix of the
input state according to the rule
\begin{equation*}
\left[
\begin{array}{cc}
\alpha _{11} & \alpha _{12} \\
\alpha _{21} & \alpha _{22}
\end{array}
\right] \rightarrow \left[
\begin{array}{cc}
I & 0 \\
0 & K^{T}
\end{array}
\right] \left[
\begin{array}{cc}
\alpha _{11} & \alpha _{12} \\
\alpha _{21} & \alpha _{22}
\end{array}
\right] \left[
\begin{array}{cc}
I & 0 \\
0 & K
\end{array}
\right] +\left[
\begin{array}{cc}
0 & 0 \\
0 & \alpha
\end{array}
\right] \equiv \alpha _{AB}.
\end{equation*}
The right hand side representing the correlation matrix of the output state
satisfies
\begin{eqnarray*}
\alpha _{AB} &\geq &\frac{i}{2}\left[
\begin{array}{cc}
I & 0 \\
0 & K^{T}
\end{array}
\right] \left[
\begin{array}{cc}
\Delta _{A} & 0 \\
0 & \Delta _{A}
\end{array}
\right] \left[
\begin{array}{cc}
I & 0 \\
0 & K
\end{array}
\right] +\frac{i}{2}\left[
\begin{array}{cc}
0 & 0 \\
0 & \pm \Delta _{B}-K^{T}\Delta _{A}K\alpha
\end{array}
\right] \\
&=&\frac{i}{2}\left[
\begin{array}{cc}
\Delta _{A} & 0 \\
0 & \pm \Delta _{B}
\end{array}
\right] ,
\end{eqnarray*}
where in the estimate of the second term we used (\ref{ppt}) with its
transpose. However, this is necessary and sufficient for the output state to
have positive partial transpose \cite{ww}. As shown in \cite{ww}, there are
nonseparable Gaussian states with positive partial transpose, therefore the
condition (\ref{ppt}) is in general weaker than (\ref{nsc}).

The condition of the theorem is automatically fulfilled in the special case
where
\begin{equation*}
K^{T}\Delta _{A}K=0.
\end{equation*}
In this case operators $R_{A}K$ commute hence $M_{A}$ sharp observable given
by is their joint spectral measure and the probability distribution $m_{\rho
}(d^{2n}z)$ can be arbitrarily sharply peaked around any point $z$ by
appropriate choice of the state $\rho .$ Hence in this case it is natural to
identify $\Phi $ as c-q (classical-quantum) channel determined by the family
of states $z\rightarrow W(z)\sigma _{B}W(z)^{\ast }$ \cite{obzor}$.$

\section{The case of one mode}\label{6}

Let us apply theorem 2 to the case of one Bosonic mode $A=B$, where
\begin{equation*}
\Delta _{A}(z,z^{\prime })=\Delta _{B}(z,z^{\prime })=\Delta (z,z^{\prime
})=x^{\prime }y-xy^{\prime }
\end{equation*}
As shown in \cite{h1}, by choosing appropriate canonical unitary
transformations $U_{1},U_{2}$, any one mode Gaussian channel with
parameters $(K,0,\alpha )$ can be transformed via
\begin{equation*}
\Phi ^{\prime }[ \rho] =U_{2}\Phi ^{\ast }\left[ U_{1}\rho
U_{1}^{\ast }\right] U_{2}^{\ast }
\end{equation*}
to one of the following normal forms, where $N\geq 0,$ $k$ is real number:
\begin{eqnarray*}
A)\quad K\left[ x,y\right]  &=&k\left[ x,0\right] ;\quad \alpha
(z,z)=\bigl(
\ N+\frac{1}{2}\bigr)\,\left( x^{2}+y^{2}\right) ; \\
B_{1})\quad K\left[ x,y\right]  &=&\left[ x,y\right] ;\quad \alpha
(z,z)=
\frac{1}{2}y^{2}; \\
B_{2})\quad K\left[ x,y\right]  &=&\left[ x,y\right] ;\quad \alpha (z,z)=\
N\left( x^{2}+y^{2}\right) ;\quad  \\
C)\quad K\left[ x,y\right]  &=&k\left[ x,y\right] ;\quad k>0,k\neq 1;\quad
\alpha (z,z)=\bigl(\ N+\frac{|1-k^{2}|}{2}\bigr)\,\left( x^{2}+y^{2}\right) ;
\\
D)\quad K\left[ x,y\right]  &=&k\left[ x,-y\right] ;\quad k>0;\quad \alpha
(z,z)=\bigl(\ N+\frac{(1+k^{2})}{2}\bigr)\,\left( x^{2}+y^{2}\right) .
\end{eqnarray*}
The case $B_{2})$ representing channel with additive classical noise, and $C)
$ representing attenuator/amplifier, are of major interest in applications
\cite{hw}, \cite{g}.

We have only to find the form $K^{T}\Delta _{A}K$ and check the
decomposability (\ref{nsc}) in each of these cases. We rely upon the simple
fact that
\begin{equation*}
\bigl(N+\frac{1}{2}\bigr)I\geq \frac{i}{2}\Delta
\end{equation*}
if and only if $N\geq 0.$

$A)$ \ \ $K^{T}\Delta K=0,$ hence $\Phi $ is c-q (in fact essentially
classical) channel;

$B)$ $\ K^{T}\Delta K=\Delta ,$ hence the necessary condition (\ref{ppt})
requires $\alpha \geq i\Delta .$ This is never fulfilled in the case $B_{1})$
due to degeneracy of $\alpha .$ Thus the channel is not entanglement
breaking (in fact it has infinite quantum capacity as shown in \cite{h1}).
On the other hand, in the case $B_{2})$ the condition (\ref{nsc}) is
fulfilled with $\nu =\mu =\alpha /2$ if and only if $N\geq 1,$ hence $\Phi $
is entanglement breaking in this case;

$C)$ $K^{T}\Delta K=k^{2}\Delta .$ It is clear that in this case the
decomposability condition holds if and only if $\alpha \geq
\frac{i}{2} (1+k^{2})\Delta ,$ which is equivalent to
$N+\frac{|1-k^{2}|}{2}\geq \frac{ (1+k^{2})}{2}$ or
\begin{equation}
N\geq \min \left( 1,k^{2}\right) .  \label{doma}
\end{equation}
This gives the condition for the entanglement breaking (which also formally
includes the case $B_{2})$);

$D)$ $K^{T}\Delta K=-k^{2}\Delta .$ Again the decomposability
condition holds if and only if $\alpha \geq
\frac{i}{2}(1+k^{2})\Delta ,$ which always holds, hence the channel
is entanglement breaking for all $ N\geq 0.$

Thus the additivity property (\ref{hiad}) holds for one-mode
Gaussian channels of the form $A),D)$ with arbitrary parameters, and
$B_{2}),C)$ with parameters satisfying (\ref{doma}). To compare this
with previous results, the only case where additivity of $C_{\chi
}(\Phi ,E,c)$ with the special energy constraint ($E=a^{\dag }a$)
was established, is $C)$ with $N=0,k<1$ (pure loss channel)
\cite{giov}, which does not intersect with our result. The actual
computation of $C_{\chi }(\Phi ,E,c)$ is in general an open problem:
there is a natural conjecture that the $\chi -$capacity of quantum
Gaussian channel with quadratic energy constraint is attained on a
Gaussian ensemble of pure Gaussian states, but so far this was only
established for c-q channels \cite{obzor} and the pure loss channel
\cite{giov}.  If the conjecture is true, then in the cases
$B_{2}),C),D)$ the optimal ensemble is the complex Gaussian
distribution $P(d^{2}z)$ with zero mean and variance $c$ on the
coherent states $W_{A}(z)\rho _{0}W_{A}(z)^{\ast }$ where $\rho
_{0}$ is the vacuum state. Hence $C(\Phi
,E,c)=g(k^{2}c+N_{0})-g(N_{0}),$ where
\begin{equation*}
N_{0}=\left\{
\begin{array}{ll}
(k^{2}-1)_{+}+N, & \mathrm{case}\, B_{2}),C); \\
k^{2}+N, & \mathrm{case}\, D).
\end{array}
\right.
\end{equation*}
is the  mean number of quanta in the output corresponding to the
vacuum input state and $g(x)=(x+1)\log (x+1)-x\log x$.

In general entanglement-breaking channels have zero quantum capacity
$Q(\Phi )=0$, cf. Sec. \ref{2}. In this connection it is notable
that the domain (\ref{doma}) coincides with zero quantum capacity
domain obtained in \cite{ww} from completely different argument.
However in any case this is superseded by the broader domain found
in \cite{g} from degradability analysis.

\textbf{Acknowledgements.} This work was partially supported by RFBR
grant 06-01-00164-a and the RAS scientific program ``Theoretical
problems of modern mathematics''. The author is grateful to  M.
Shirokov, V. Giovannetti and M. Wolf for discussions.

\end{document}